\documentclass{article}

\usepackage{graphicx}
\usepackage{subfigure}
\usepackage{amsmath}
\usepackage{xspace}

\newcommand{\eepp}{\ensuremath{e^+e^-\to\pi^+\pi^-}\xspace}
\newcommand{\eeee}{\ensuremath{e^+e^-\to e^+e^-}\xspace}
\newcommand{\eemm}{\ensuremath{e^+e^-\to\mu^+\mu^-}\xspace}

\newcommand{\eeeeg}{\ensuremath{e^+e^-\to e^+e^-(\gamma)}\xspace}
\newcommand{\ee}{\ensuremath{e^+e^-}}
\newcommand{\mm}{\ensuremath{\mu^+\mu^-}\xspace}
\newcommand{\mumu}{\ensuremath{\mu^+\mu^-}\xspace}

\newcommand{\mmg}{\ensuremath{\mu^+\mu^-(\gamma)}\xspace}
\newcommand{\ppg}{\ensuremath{\pi^+\pi^-(\gamma)}\xspace}

\newcommand{\BW}{\mathrm{BW}}
\newcommand{\GS}{\mathrm{GS}}
\newcommand{\mrho}{\mathrm{M}_\rho}
\newcommand{\momg}{\mathrm{M}_\omega}

\title{High-statistics measurement of the pion form factor in the
  $\rho$-meson energy range with the CMD-2 detector}

\author{
R.\,R.\,Akhmetshin$^{a}$,
V.\,M.\,Aulchenko$^{a,b}$,
V.\,Sh.\,Banzarov$^{a}$,
L.\,M.\,Barkov$^{a,b}$,  \and
N.\,S.\,Bashtovoy$^{a}$,
A.\,E.\,Bondar$^{a,b}$,
D.\,V.\,Bondarev$^{a,b}$,
A.\,V.\,Bragin$^{a}$,  \and
S.\,K.\,Dhawan$^{d}$,
S.\,I.\,Eidelman$^{a,b}$,
D.\,A.\,Epifanov$^{a}$,
G.\,V.\,Fedotovich$^{a,b}$, \and
N.\,I.\,Gabyshev$^{a}$,
D.\,A.\,Gorbachev$^{a}$,
A.\,A.\,Grebenuk$^{a}$,
D.\,N.\,Grigoriev$^{a,b}$,  \and
\fbox{V.\,W.\,Hughes$^{d}$},
F.\,V.\,Ignatov$^{a}$,
S.\,V.\,Karpov$^{a}$,
V.\,F.\,Kazanin$^{a,b}$, \and
B.\,I.\,Khazin$^{a,b}$,
I.\,A.\,Koop$^{a,b}$,
P.\,P.\,Krokovny$^{a,b}$,
A.\,S.\,Kuzmin$^{a,b}$, \and
I.\,B.\,Logashenko$^{a,c}$,
P.\,A.\,Lukin$^{a,b}$,
A.\,P.\,Lysenko$^{a}$,
K.\,Yu.\,Mikhailov$^{a}$, \and
J.\,P.\,Miller$^{c}$,
A.\,I.\,Milshtein$^{a,b}$,
I.\,N.\,Nesterenko$^{a,b}$,
M.\,A.\,Nikulin$^{a}$, \and
V.\,S.\,Okhapkin$^{a}$,
A.\,V.\,Otboev$^{a}$,
E.\,A.\,Perevedentsev$^{a,b}$,
A.\,S.\,Popov$^{a}$,\and
S.\,I.\,Redin$^{a}$, 
B.\,L.\,Roberts$^{c}$,
N.\,I.\,Root$^{a}$,
A.\,A.\,Ruban$^{a}$,
N.\,M.\,Ryskulov$^{a}$,\and
A.\,G.\,Shamov$^{a}$,
Yu.\,M.\,Shatunov$^{a}$,
B.\,A.\,Shwartz$^{a,b}$,
A.\,L.\,Sibidanov$^{a}$,\and
V.\,A.\,Sidorov$^{a}$,
A.\,N.\,Skrinsky$^{a}$,
V.\,P.\,Smakhtin$^{f}$,
I.\,G.\,Snopkov$^{a}$,\and
E.\,P.\,Solodov$^{a,b}$,
\fbox{J.\,A.\,Thompson$^{e}$},
Yu.\,V.\,Yudin$^{a}$,
A.\,S.\,Zaitsev$^{a,b}$, \and
S.\,G.\,Zverev$^{a}$
}

\date{}
\begin{document}
\maketitle
\begin{center}
  {\it $^a$Budker Institute of Nuclear Physics, 630090,
    Novosibirsk, Russia} \\
  {\it $^b$Novosibirsk State University, 630090,
    Novosibirsk, Russia} \\
  {\it $^c$Boston University, Boston, MA 02215, USA}\\
  {\it $^d$Yale University, New Haven, CT 06511, USA}\\
  {\it $^e$University of Pittsburgh, Pittsburgh, PA 15260, USA}\\
  {\it $^f$Weizmann Institute of Science,  76100, Rehovot, Israel}\\
\end{center}

\maketitle

\begin{abstract}
We present a measurement of the pion form factor based on $e^+e^-$
annihilation data from the CMD-2 detector in
the energy range $0.6<\sqrt{s}<1.0$ GeV with a systematic uncertainty
of 0.8\%. A data sample is five times larger
than that used in our previous measurement.
\end{abstract}

\section{Introduction}

Measurement of the pion form factor $|F(q^2)|$, or the cross section
$\eepp$, in the center-of-mass (c.m.) energy range $\sqrt{s}<1.4$ GeV 
is interesting for studies of the pion internal structure,  
for the analysis of the properties of the 
light vector mesons ($\rho$,~$\omega$,~$\varphi$) and their excitations,
and as a determination of an important contribution to $R(s)$. 

The total cross section $\ee\to hadrons$, which is dominated by the $\eepp$
cross section in the energy range under discussion, is
often expressed as the
dimensionless ratio 
\begin{equation}
\label{eq:R}
R(s) = \frac{\sigma(\ee\to hadrons)}{\sigma(\ee\to\mumu)},
\end{equation}
and has been an important topic in high energy physics since the quark
model was established. Recent interest to the measurement of
$R(s)$ was stimulated by the measurement of the anomalous magnetic 
moment of the muon $a_\mu$ at BNL \cite{E821} with the unprecedented 
precision of 0.54  ppm. The
 measured value of $a_\mu$ is about 2 to 3 standard deviations above the
 Standard Model expectation, which could indicate the long-sought
 existence of New Physics.
When integrated with the proper kernel function, $R(s)$ gives a value
 of $a^{(\rm had,LO)}_{\mu}$ --- the leading order hadronic contribution to
 $a_\mu$. The accuracy of the SM prediction, currently $\approx 0.55$
 ppm \cite{DEHZ}, is dominated by the 
knowledge of $R(s)$ at $\sqrt{s}<2$ GeV, where the measurement of
the $\ee\to hadrons$ cross section is the only source of $R(s)$.

The energy range $0.36<\sqrt{s}<1.4$ GeV has been studied at the
electron-positron collider VEPP-2M \cite{VEPP} (Novosibirsk, Russia). Two
experiments, CMD-2 \cite{CMD2,PTE} and SND \cite{SND}, started in 1992 and
1995, respectively, and continued up to 2000, when the collider was
shut down. The CMD-2 detector consists of the drift chamber, the
proportional Z-chamber, the barrel CsI calorimeter, the endcap BGO
calorimeter and the muon range system. The
drift chamber, Z-chamber and the endcap calorimeters are placed
inside a thin superconducting solenoid with a field of 1 T.

CMD-2 collected $e^+e^-\to\pi^+\pi^-$ data at five separate energy
scans, starting from 1994. Approximately
$10^6$ \eepp events were selected for analysis. From analysis
perspectives, the VEPP-2M energy range is naturally subdivided into
three intervals. In the energy range $0.36<\sqrt{s}<0.6$ GeV,
covered in the 1996 data taking run, the momentum resolution of the drift
chamber is good enough to separate the $e$, $\mu$ and $\pi$ in the
final states \cite{Lesha}. At  higher energies, the energy
deposition in the calorimeter is used for the separation. The
energy range  $1.0<\sqrt{s}<1.4$ GeV, covered in the 1997 run
\cite{fedya}, is distinguished by the relatively small value of the  
$\eepp$ cross section. The bulk of the data were
collected in the energy range $0.6<\sqrt{s}<1.0$ GeV, where the
cross section is enhanced by the $\rho(770)$ resonance. This interval was
covered in the 1994-95 run \cite{rhoart,rhoerrata}, and later the
measurement was repeated in 1998 with five times larger 
integrated luminosity. The analysis of the 1998 data sample
is discussed in this paper. 

\section{Data Analysis}

\subsection{Overview}

The data were collected at 29 energy points covering the c.m. energy range 
$0.6<\sqrt{s}<0.98$ GeV with energy steps varying from 0.001 GeV at the 
$\omega$-meson peak to 0.030 GeV at the $\rho$-resonance tails.

The analysis follows the approach used in our previous
measurement \cite{rhoart,rhoerrata}.
From $1.5\cdot 10^8$ triggers recorded, about $2.6\cdot 10^6$ events were 
selected as \textit{collinear}, with a signature of two particles of 
opposite charges and nearly back-to-back momenta originating from the 
interaction point. The following selection criteria were used:
\begin{enumerate}
\item The event was triggered by the trackfinder.
\item Two tracks of opposite  charge originating from the interaction 
  region were reconstructed in the drift chamber.
\item Each track was reconstructed using at least 7 wire hits.
\item The minimal distance from two tracks to the beam axis, $\rho$,
  is less than 0.3 cm and z-coordinate of the vertex (along the
  beam axis) is within $-15<z<15$ cm.
\item The average momentum of the two particles $(p_1+p_2)/2$ is 
  between 200 and 600 MeV/c.
\item The transverse momentum of each track is above 100 MeV/c.
\item The difference between the azimuthal angles (in the plane
  perpendicular to the beam axis) of the two particles
  $ |\Delta \varphi |=|\pi -|\varphi _{1}-\varphi _{2}||<0.15$. 
\item The difference between the polar angles (the angle between 
  the momentum and the beam axis) of the two particles
  $|\Delta \Theta |=|\Theta _{1}-(\pi -\Theta _{2})|<0.25$.
\item The average polar angle of the two particles $ \Theta_{avr}=[\Theta
  _{1}+(\pi -\Theta _{2})]/2 $ is within $1.1<\Theta_{avr}<(\pi-1.1)$.
  This criterion determines the fiducial volume.
\end{enumerate}

The sample selected contains beam-originating \eeee, \eemm, \eepp
 events and the small background of cosmic particles, mostly muons,
 which pass near the
 interaction region and mimic collinear events. The number of the
 background events $N_{cosmic}$ is determined from the analysis of the
 spatial distribution of the vertex. 

The three beam-originating final states were separated using the
information on the particle energy deposition in the CsI 
calorimeter. The separation procedure is based on the minimization of
the unbinned likelihood function:
\begin{equation}
\label{lfunc1}
L= - \sum_{events} \ln \left( \sum_a N_a\cdot f_a(E^+,E^-) \right) +
\sum_a N_a,
\end{equation}
where $a$ is the final state ($a=ee$, $\mu\mu$, $\pi\pi$, $cosmic$),
$N_a$ is the number of events of the type $a$ and $f_a(E^+,E^-)$
is the probability density function (p.d.f.) for an event of type $a$ 
to have the energy depositions $E^+$ and $E^-$. 

It is assumed that $E^+$ 
and $E^-$ are independent for events of the same type,
therefore the p.d.f. can be factorized as
\[ f_a(E^+,E^-)=f_a^+(E^+)\cdot f^-_a(E^-), \]
where $f^{\pm}_a(E)$ are the energy deposition p.d.f.s for individual 
$e^\pm$, $\mu^\pm$, $\pi^\pm$ and cosmic muons.
This assumption is not entirely correct.
The energy deposition depends on the calorimeter thickness seen by particle
($\approx 8X_0$ at $90^0$). 
Since the incident angles at which the two particles in the final
state hit the calorimeter are nearly the same,
that leads to a correlation between $E^+$ and $E^-$. 
This effect is corrected for by the recalibration of the
energy deposition.
The second source of the correlation is introduced by initial
state radiation. This effect will be discussed in more detail in
section\ \ref{sec:other}.

The overlap between the energy deposition of electrons and pions is rather
small, which makes the described procedure very robust. However, the
energy deposition of muons and pions is not that different, therefore
the small errors in the p.d.f.\ for these
particles lead to a large correlated error for $N_{\pi\pi}$ and
$N_{\mu\mu}$. To avoid this problem, the ratio of the number of 
$\mu^+\mu^-$ pairs to the number of $e^+e^-$ pairs is fixed during the
minimization at the value calculated according to QED with
radiative corrections and detection efficiencies taken into account:
\[
\frac{N_{\mu\mu}}{N_{ee}}=
\frac{\sigma_{\mu\mu}\cdot (1+\delta_{\mu\mu}) \varepsilon_{\mu\mu}}
{\sigma_{ee}\cdot (1+\delta_{ee}) \varepsilon_{ee}},
\]
where $\sigma$ are the Born cross-sections, $\delta$ are the
radiative corrections
and \( \varepsilon  \) are the efficiencies.

The specific form of the energy deposition functions (p.d.f.s) was
evaluated in the variety of studies. P.d.f.s for electrons
(positrons) and background muons were obtained with the specially
selected subsets of data. P.d.f.s for minimum ionizing particles (muons
and pions without nuclear interactions) were extracted from the
simulation. The complete p.d.f.\ for pions with nuclear interactions
taken into account was obtained from the analysis of the energy deposition
of tagged pions coming from the $\phi(1020)\to\pi^+\pi^-\pi^0$ decay,
a high-statistics measurement of which was performed at CMD-2 
in separate data taking runs \cite{3pi}. In all cases, only the
functional form of p.d.f.s was fixed. The
particular values of the function parameters were determined by the
minimization procedure.

To simplify the final error calculation, the 
likelihood function (\ref{lfunc1}) 
is rewritten to have the following global fit parameters:
$(N_{ee}+N_{\mu\mu})$ and $N_{\pi\pi }/(N_{ee}+N_{\mu\mu})$ 
instead of $N_{ee}$ and  $N_{\pi\pi}$ (with $N_{\mu\mu}/N_{ee}$ and
$N_{cosmic}$ fixed).

The pion form factor is calculated as:
\begin{equation}
\label{piform}
|F_\pi|^2 = \frac{N_{\pi\pi}}{N_{ee}+N_{\mu\mu}} \times 
\frac{
\sigma_{ee}\cdot(1+\delta_{ee})\varepsilon_{ee} +
\sigma_{\mu\mu}\cdot(1+\delta_{\mu\mu})\varepsilon_{\mu\mu}
}{\sigma_{\pi\pi}\cdot(1+\delta_{\pi\pi})
(1+\Delta_N)(1+\Delta_D)\varepsilon_{\pi\pi}\cdot(1+\Delta_{sep})}
-\Delta_{3\pi},
\end{equation}
where the ratio $N_{\pi\pi}/(N_{ee}+N_{\mu\mu})$ is determined in the
minimization of  (\ref{lfunc1}),
$\sigma$ are the corresponding Born cross sections, integrated over
the fiducial volume,
$\delta$ are the radiative corrections, $\epsilon$ are the detection 
efficiencies, 
$\Delta_D$ and $\Delta_N$ are the corrections for the pion losses
caused by decays in flight and nuclear interactions, respectively,
$\Delta_{3\pi}$  is the correction for $\omega\to\pi^+\pi^-\pi^0$
background and $\Delta_{sep}$ is the correction for the systematic
shift, introduced by the separation procedure.
In the case of \eepp, $\sigma_{\pi\pi}$ corresponds to point-like pions.

\subsection{\label{sec:eff}Efficiency}

The most significant difference between 1994-95 and 1998 data analyses
lies in the measurement of the detection efficiency $\varepsilon$.
The efficiency $\varepsilon$ is the product of the reconstruction
efficiency and the trigger efficiency.

For the 1994-95 data the reconstruction efficiency is high 
($\approx$ 97\%--99\%) and the same for all three final
states --- studies showed that the difference does not exceed
0.2\%. Therefore, the efficiencies cancel in
\eqref{piform}. 

In the run of 1998 the CMD-2 drift chamber showed signs of
aging. That led to a lower reconstruction efficiency and, 
more important, to different values of the efficiency for the three final
states. This difference, if unaccounted for, would lead to a
significant systematic error on the form factor. 
Therefore, a  direct measurement of the
reconstruction efficiency for all types of collinear events was
necessary.

This measurement is based on a well-known
technique. A test sample of collinear events is selected using
criteria based on the calorimeter data, which are uncorrelated with
the standard selection criteria based on the information from the
tracking system. The efficiency is calculated as a fraction
of test events, which passed the standard selection criteria 
for collinear events.

The selection criteria for the test sample are the following:
\begin{enumerate}
\item The event was triggered by the trackfinder. 
\item There are exactly two clusters in the calorimeter.
\item There is a hit in the Z-chamber near each cluster. This
  requirement selects the clusters produced by a charged particle.
\item The clusters are collinear if one takes into account the particle
  deflection in the detector magnetic field: 
\begin{eqnarray*}
  & |\pi -(\Theta _{1}+\Theta _{2})|<0.1, & \\
  & \left| \left| \pi -|\varphi _{1}-\varphi _{2}|\right| -
  \varphi_0 \right| <0.1, &
\end{eqnarray*}
where \( \Theta  \) and \( \varphi  \) are the polar and azimuthal
angles of the cluster and $\varphi_0$ is the expected azimuthal
deflection angle 
of particles in the CMD-2 magnetic field of 1 T.
\end{enumerate}

\begin{figure}
\begin{center}
\includegraphics[width=0.49\textwidth]{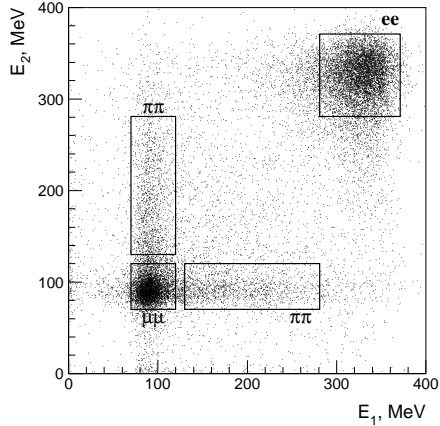}
\end{center}
\caption{\label{fig:energycut_all} Definition of three classes of
  test events}
\end{figure}

The test event sample is subdivided into three classes:
\begin{enumerate}
\item For the \eeee subset, the energy deposition of each cluster is between \(
  E_{min}=(0.82\cdot E_{B}-40) \) and \( E_{max}=(0.82\cdot E_{B}+50) \) MeV.
\item For the \eepp subset, the energy deposition of one 
  cluster is between 70 and 120 MeV and the energy deposition of
  another cluster is between 120 MeV and $E_{min}=(0.82\cdot
  E_{B}-40)$ MeV.
\item For the \eemm subset, the energy deposition of each
  cluster is between 70 and 120 MeV.
\end{enumerate}
The definitions of the three classes of test events are demonstrated in
Fig.\ \ref{fig:energycut_all}. 

\begin{figure}
\begin{center}
\begin{tabular}{cc}
\subfigure[\label{fig:mmgood} Test ``$\mu\mu$''
  events which passed standard selection criteria]
{\includegraphics[width=0.49\textwidth]{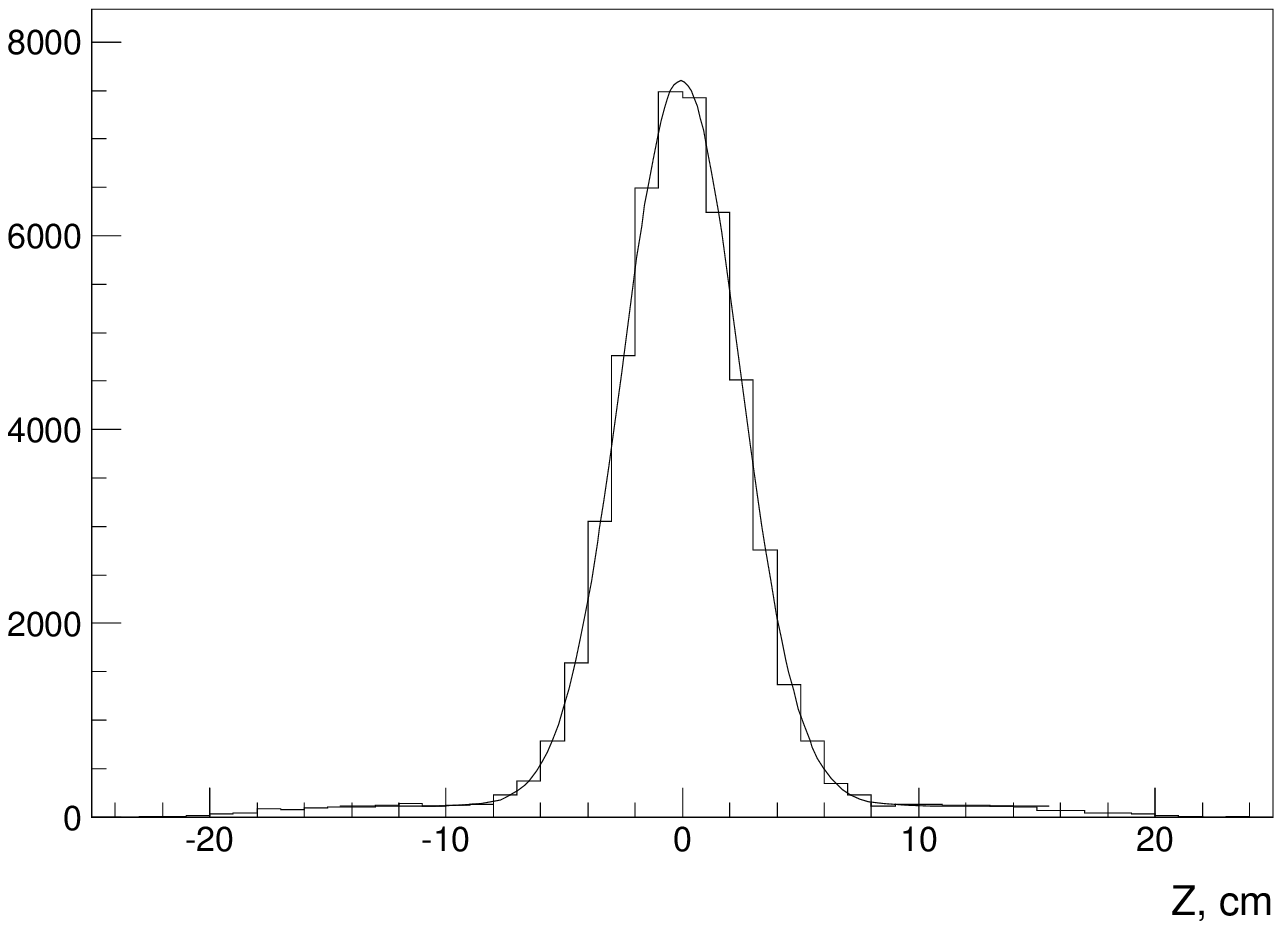}} &
\subfigure[\label{fig:mmbad} Test ``$\mu\mu$''
  events which failed standard selection criteria]
{\includegraphics[width=0.49\textwidth]{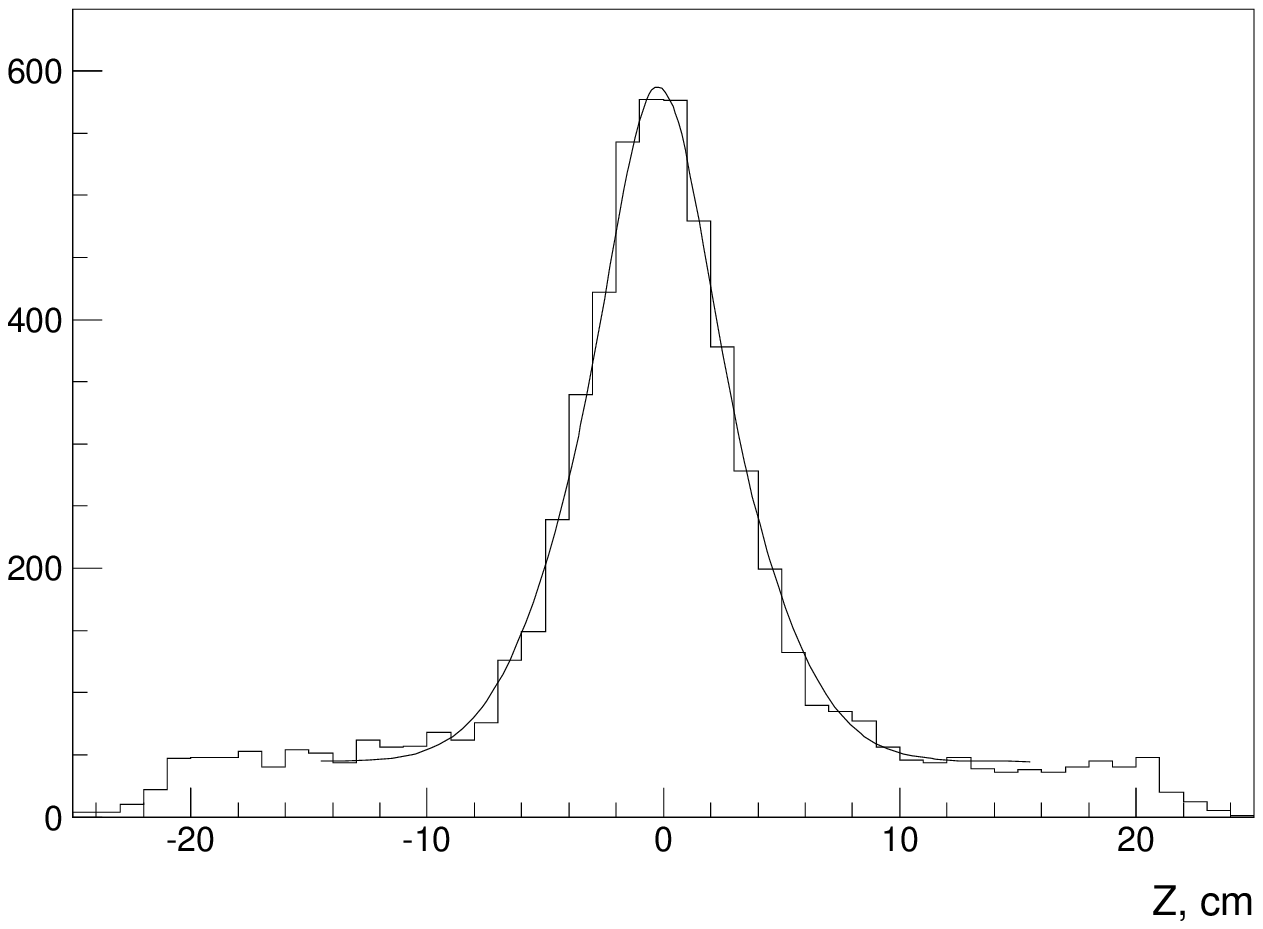}}
\end{tabular}
\end{center}
\caption{\label{fig:effz} Position of the track origin along the beam
  axis. The distributions are the sum of the gaussian-like signal and
  the flat background.}
\end{figure}

\begin{figure}
\begin{center}
\begin{tabular}{cc}
\subfigure[\label{fig:effnee} Reconstruction efficiency for \eeee events]
{\includegraphics[width=0.46\textwidth]{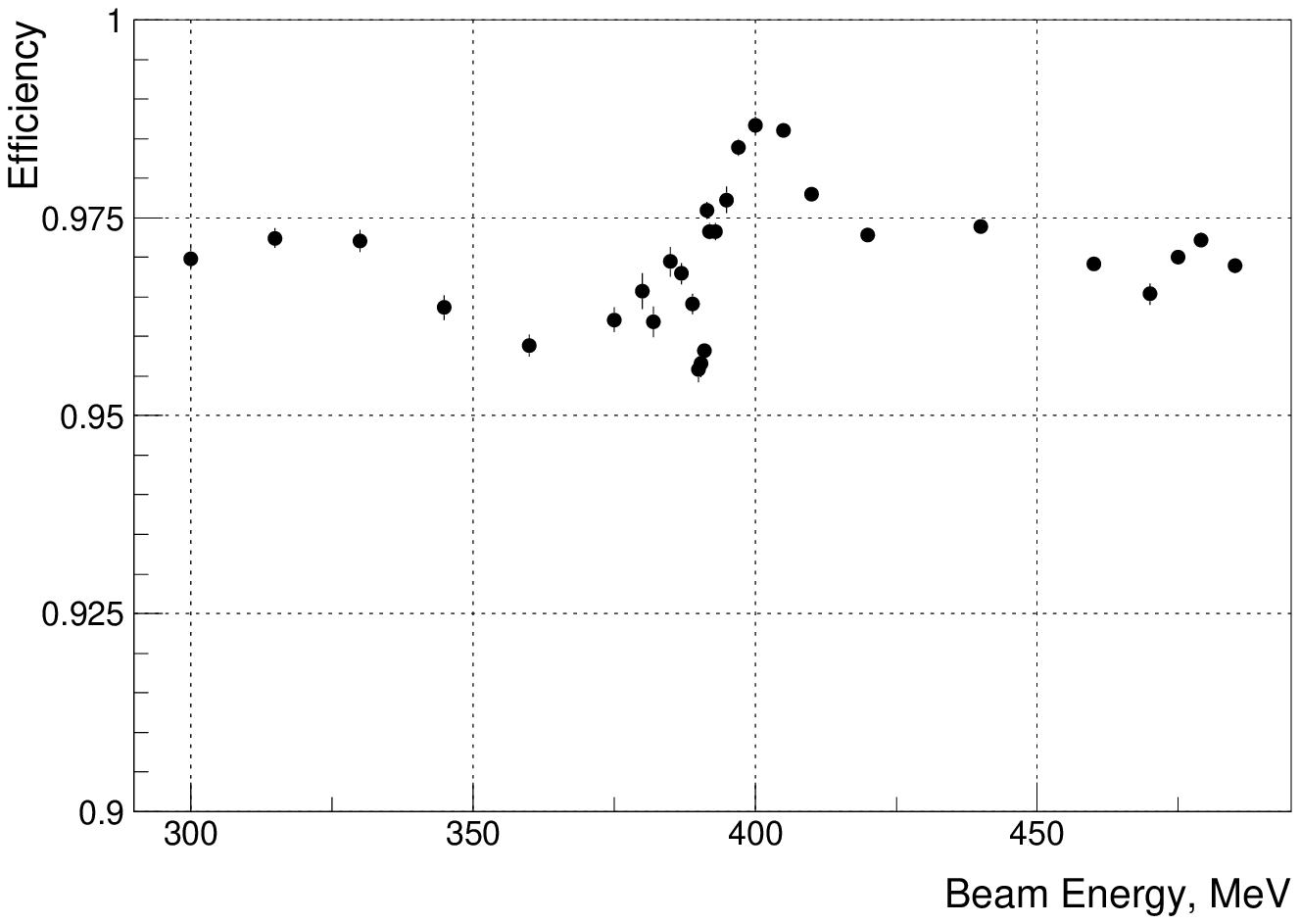}} &
\subfigure[\label{fig:effnmipee} Ratio of combined reconstruction
  efficiency of \eemm and \eepp events and efficiency of \eeee events]
{\includegraphics[width=0.46\textwidth]{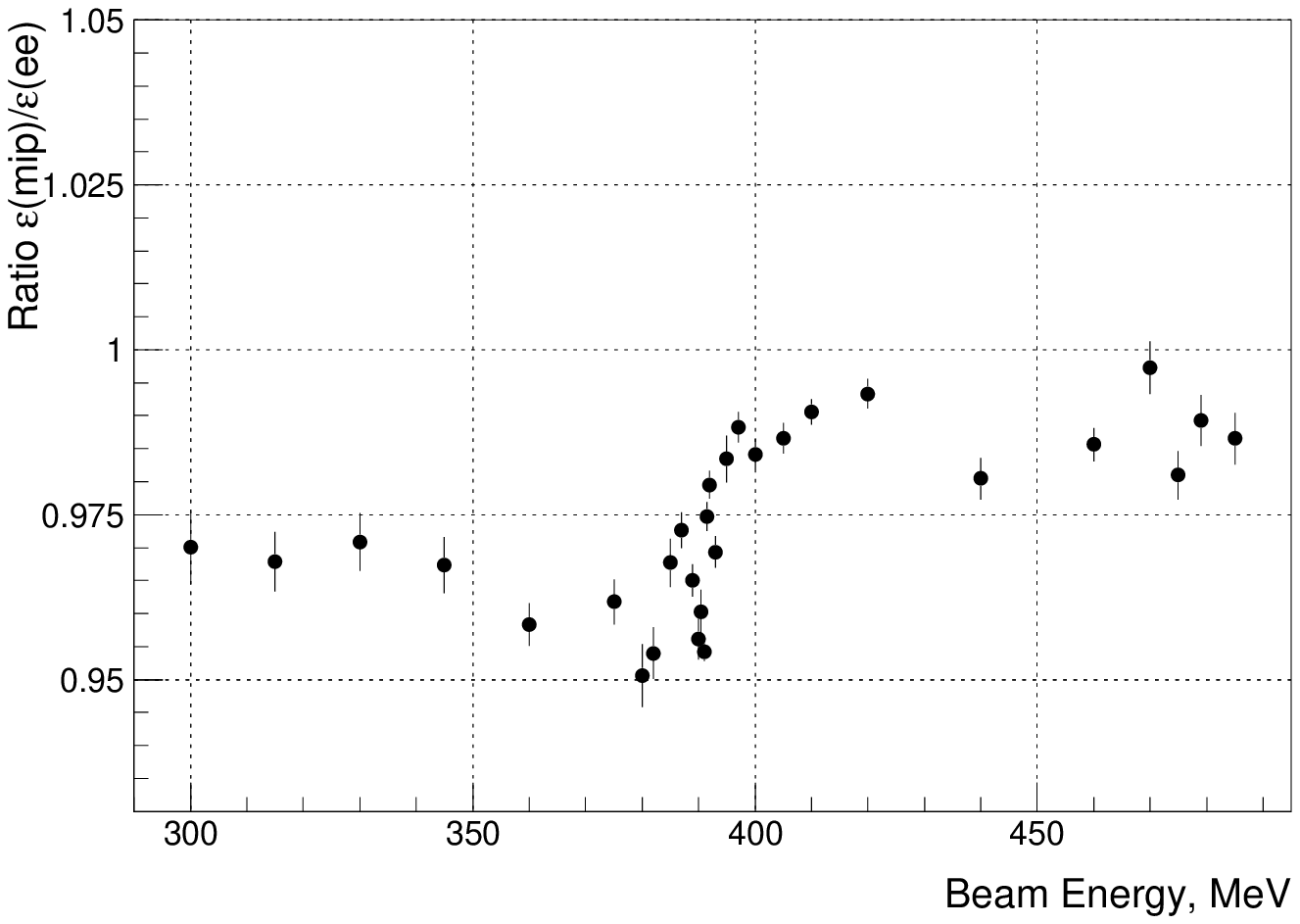}}
\end{tabular}
\end{center}
\caption{\label{fig:effall} Measurement of the reconstruction efficiency
  for all types of collinear events}
\end{figure}

The \eeee events have a unique signature of two high-energy clusters
in the calorimeter, therefore this subset contains virtually no background. 
On the contrary, the \eepp and \eemm test samples contain a significant
contribution of cosmic muon  background. 
To subtract the background, 
the additional requirement to have at least one reconstructed track was
added to the selection criteria for test events. 
This cut rejects only $\approx 0.1\%$ of
events, and therefore does not introduce a significant contribution to
the systematic error of the efficiency measurement. For each
class of the test events, the distributions of the z-coordinate of the
track origin were collected for events which pass the standard selection
criteria and for events which failed them. The distributions, shown in
Fig.\ \ref{fig:effz}, were fitted with the  combination of a
Gaussian-like distribution, which represents the beam-originating
events, and a uniform distribution, which represents the background
events. The efficiency for a particular class of test events is
calculated as 
\[
\varepsilon = \frac{N_{\text{pass}}}{N_{\text{pass}}+N_{\text{fail}}},
\]
where $N_{\text{pass}}$ and $N_{\text{fail}}$, obtained from the fit,
are the integrals of the Gaussian-like distributions for the cases,
when test event pass and fail the standard selection criteria. 
The results of the efficiency measurement, $\varepsilon_{ee}$ and
$\varepsilon_{\pi\pi,\mu\mu}/\varepsilon_{ee}$, are shown in Fig.\
\ref{fig:effall}. The wave-like structure of the efficiency as a
function of energy is explained by varying conditions of data 
taking. The drift chamber performance was changing during the run,
generally degrading with time as we scanned from higher to lower
energies. Additional problems occurred during a long period
when data were taken around the $\omega$ meson mass.

\begin{figure}
\begin{center}
\begin{tabular}{cc}
\subfigure[\label{fig:effsim_rat} Ratio of the reconstruction
  efficiencies for \eepp and \mm events separately and the combined efficiency]
{\includegraphics[width=0.49\textwidth]{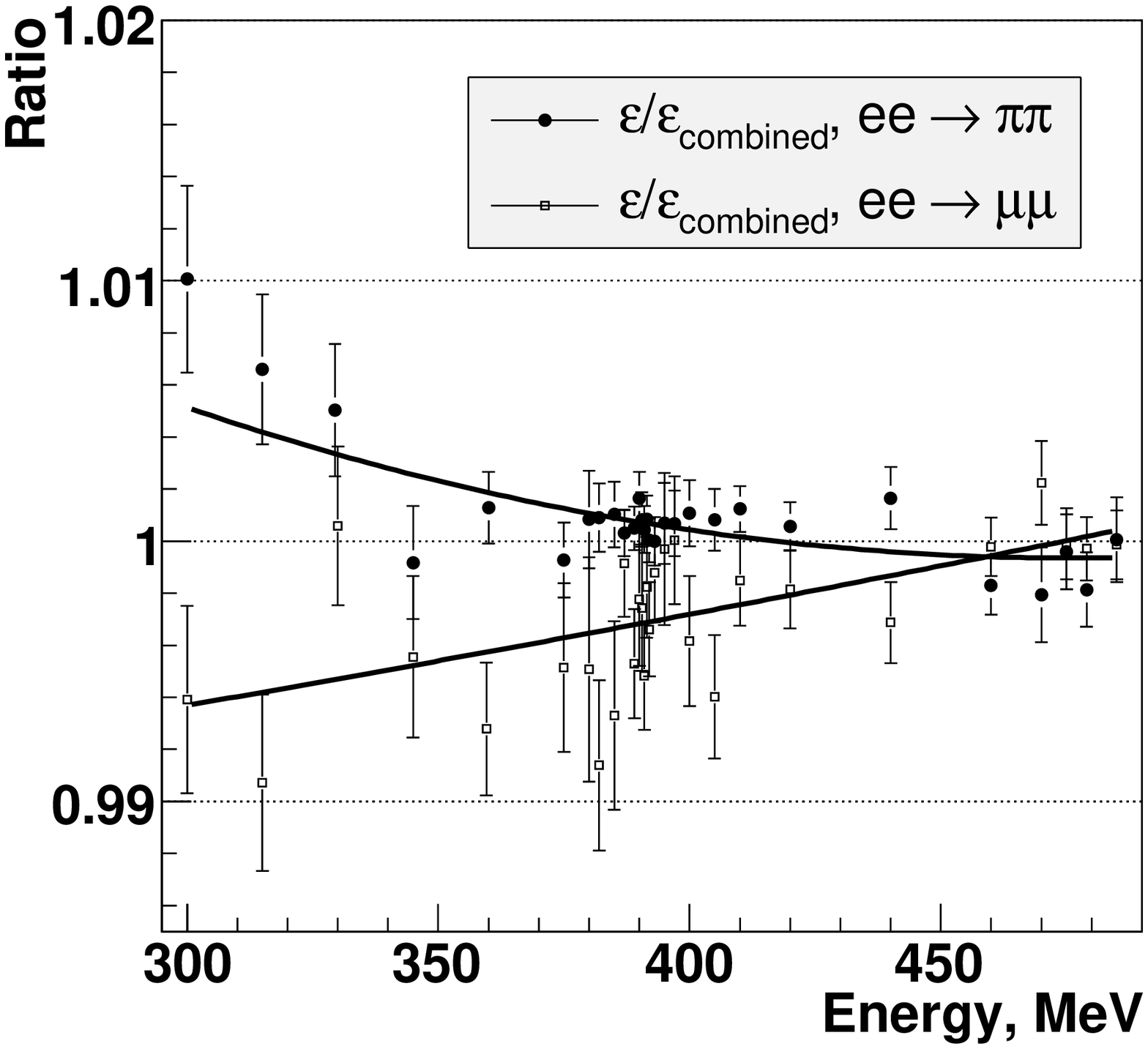}} &
\subfigure[\label{fig:effsim_brem} Contribution to the efficiency for
  \eeee events from the bremsstrahlung in the beam pipe material]
{\includegraphics[width=0.49\textwidth]{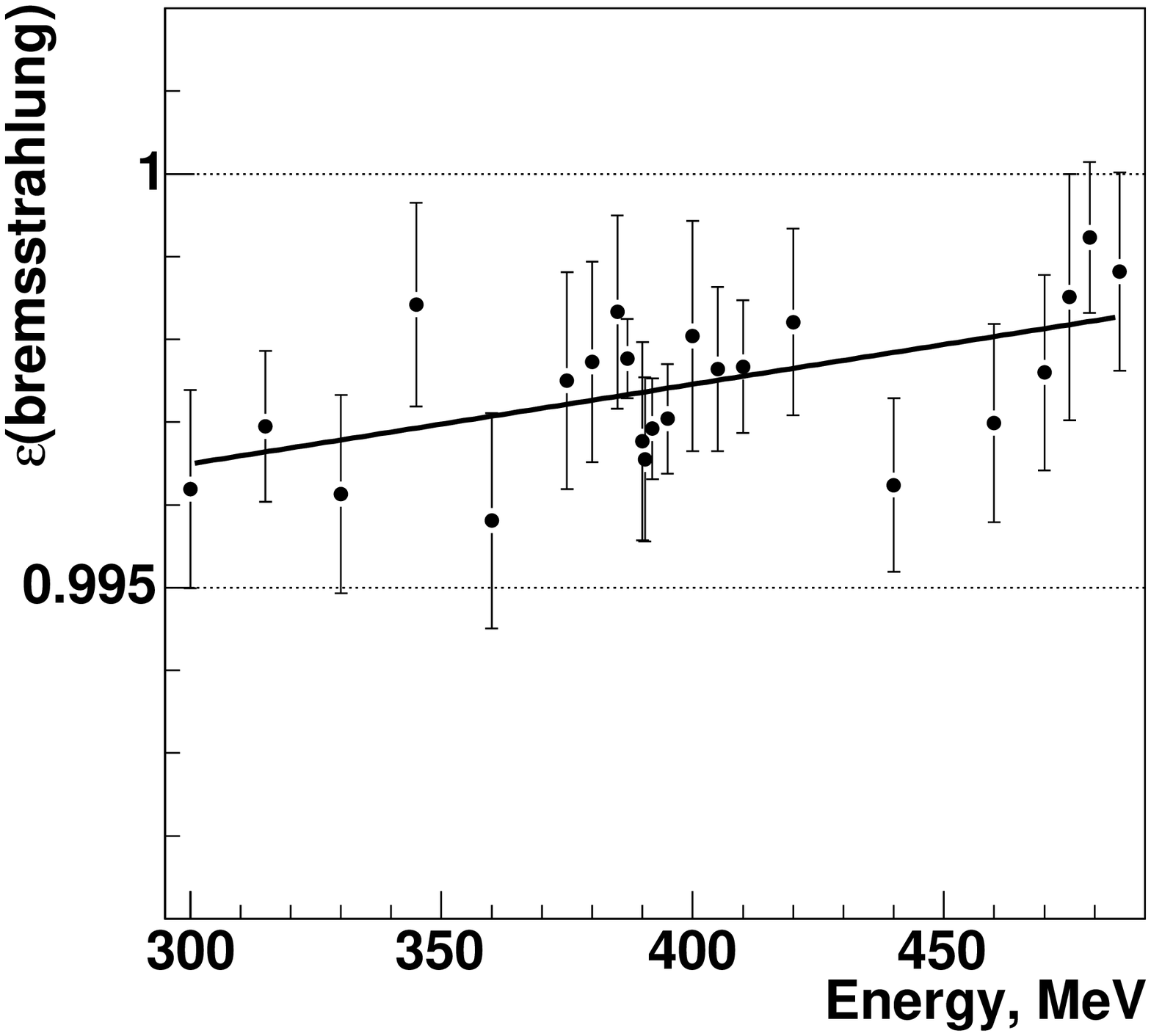}}
\end{tabular}
\end{center}
\caption{\label{fig:effsim} Reconstruction efficiency measurement for
  the simulated data set}
\end{figure}

The \eemm test sample in addition to \eemm contains those \eepp
events, where both pions in the final state interact as the
minimum-ionizing particles (MIP). While it is possible to 
extract separate efficiencies for \eemm and \eepp events, the results
have poor statistical precision. Therefore, a different approach was
used. Since the difference between $\mu$- and $\pi$-signals in
the drift chamber is much smaller than the difference between $\mu$-
and $e$-signals, the difference
$(\varepsilon_{\mu\mu}-\varepsilon_{\pi\pi})$ is much smaller than
$(\varepsilon_{ee}-\varepsilon_{\pi\pi})$. 
That allows us to measure the combined efficiency $\varepsilon_{MIP}$
using the combined \eemm and \eepp test event samples, and then to
extract the individual efficiencies $\varepsilon_{\mu\mu}$ and
$\varepsilon_{\pi\pi}$, applying a small correction estimated with the
help of the GEANT simulation (Fig.\ \ref{fig:effsim_rat}) 
to $\varepsilon_{MIP}$.

The procedure described
does not account for Bhabha events where the electron or
positron in the final state radiates a high-energy photon while
passing the wall of the 
beam pipe or the inner part of the drift chamber. Such events mostly ($>$90\%)
disappear from the test sample, as they typically have more than two
clusters in the calorimeter. This contribution to the Bhabha
reconstruction inefficiency was evaluated in the separate
simulation. It changes slowly from 0.3\% at $\sqrt{s}=600$ MeV to 0.2\%
at $\sqrt{s}=1000$ MeV (Fig.\ \ref{fig:effsim_brem}).

The efficiency measurement was tested with the full GEANT simulation,
where the \eeeeg, \mmg, \ppg events were generated and mixed
together. The drift chamber performance in the simulation was tuned to
represent adequately the performance seen with the data.
The complete procedure described above was applied to
the simulated data. It was found that the
difference between the measured and the true efficiencies does not
exceed 0.2\%.

\begin{figure}
\begin{center}
\begin{tabular}{cc}
\subfigure[\label{fig:effb_ee} Trigger efficiency for \eeee events]
{\includegraphics[width=0.46\textwidth]{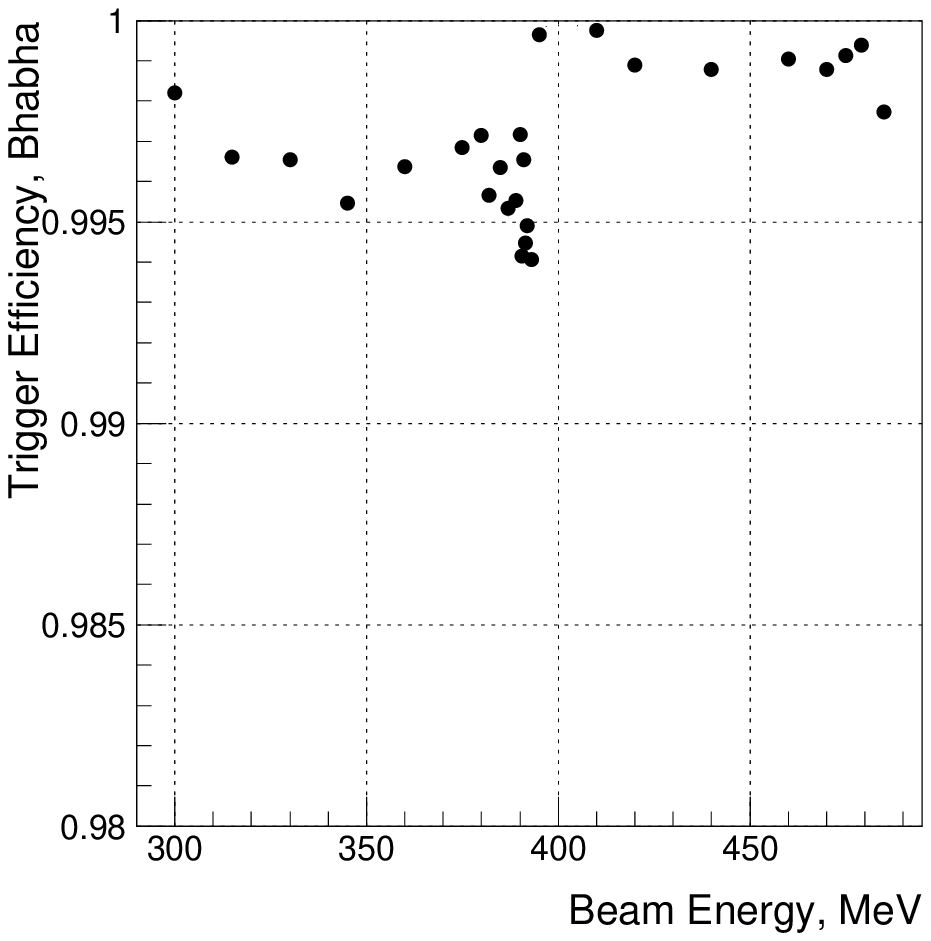}} &
\subfigure[\label{fig:effb_ratio} Ratio of the trigger
  efficiency for \eemm and \eepp events and for \eeee events]
{\includegraphics[width=0.46\textwidth]{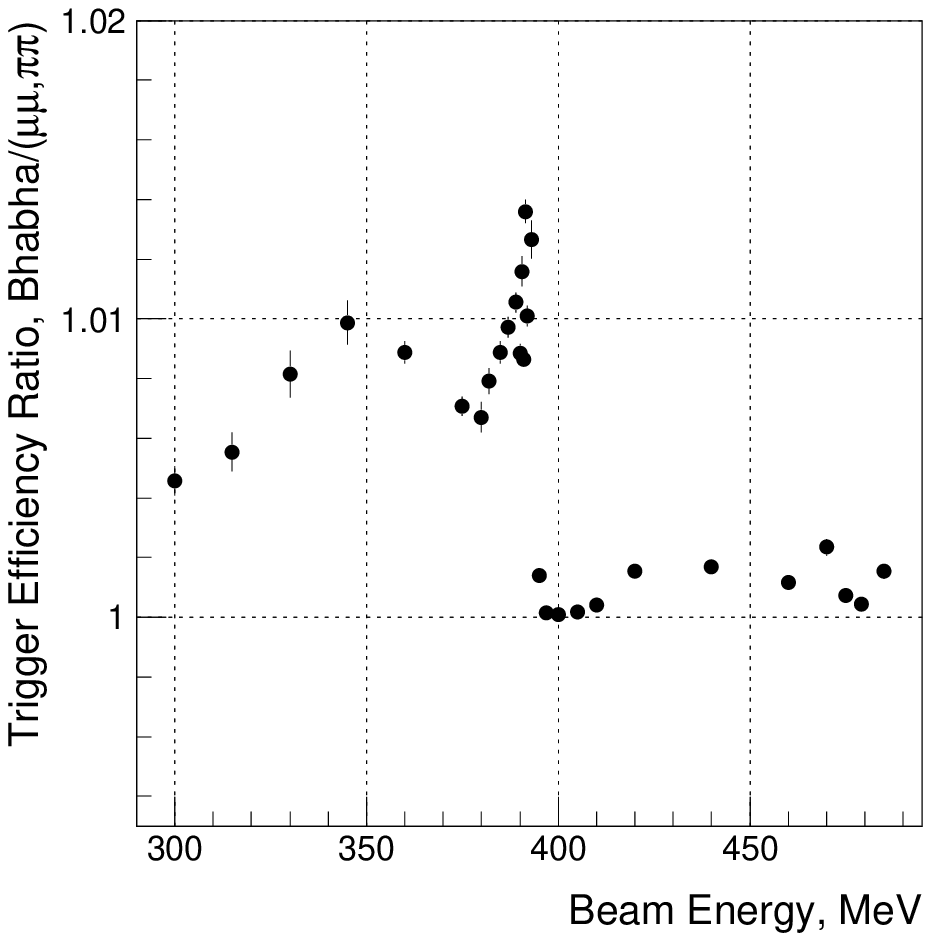}} 
\end{tabular}
\end{center}
\caption{\label{fig:effb} Measurement of the trigger efficiency}
\end{figure}

The events used in the form factor analysis were triggered by the
trackfinder  --- 
the dedicated track processor. The trackfinder generated positive
decision when at least one track candidate was identified in the
event. Information about all identified candidates is saved in the raw
data stream.

The trigger efficiency measurement is based on the fact that there are
two well-separated tracks in the final state.
Using one track to ensure the trigger, the efficiency for a single
track was calculated as the probability for the trackfinder to
identify a track candidate in the vicinity of the second
track. The trigger efficiency $\varepsilon_{t}$ was calculated from the
single track efficiency $\varepsilon_1$ as 
$\varepsilon_{t}=2\varepsilon_1-\varepsilon_1^2$.

The same three samples of test events as for the reconstruction
efficiency were used to measure the trigger efficiency. The
contribution of the cosmic
background was subtracted using the $Z$-distribution of
the track origin. The results of the measurement are shown in Fig.\
\ref{fig:effb}. The trigger efficiency for \eemm and \eepp was found
to be the same within 0.2\%. The difference between efficiencies for
\eeee and \eepp events, important for the form factor analysis,
is negligible for energies $\sqrt{s}\geq 0.79$ GeV. For the lower
energies the difference increases to $\approx 0.5-1\%$, which
coincides in time with the malfunction of one of the elements of the
tracking system. 

\subsection{\label{sec:rc}Radiative corrections}

The radiative corrections $\delta$ in \eqref{piform} were calculated
according to 
\cite{rc3}. The radiative corrections for \eeee and \eemm 
account for the radiation by the initial and final
particles and for the effects of the vacuum polarisation. The
radiative corrections for 
\eepp account for only the radiation by the initial and final
particles. 

The calculation was performed using the fast Monte Carlo
technique. The events have been first generated with the weak cuts in
the wide solid angle, then the
angles of the particles were smeared with the detector resolution and,
finally, the selection cuts were applied. The detector resolution was
obtained from the fit of the experimental $\Delta\varphi$ and
$\Delta\Theta$ distributions with the
convolution of the ideal $\Delta\varphi$ and $\Delta\Theta$
distributions, obtained from the primary generator, and the
detector response function.

\begin{figure}
\begin{center}
\includegraphics[width=\textwidth]{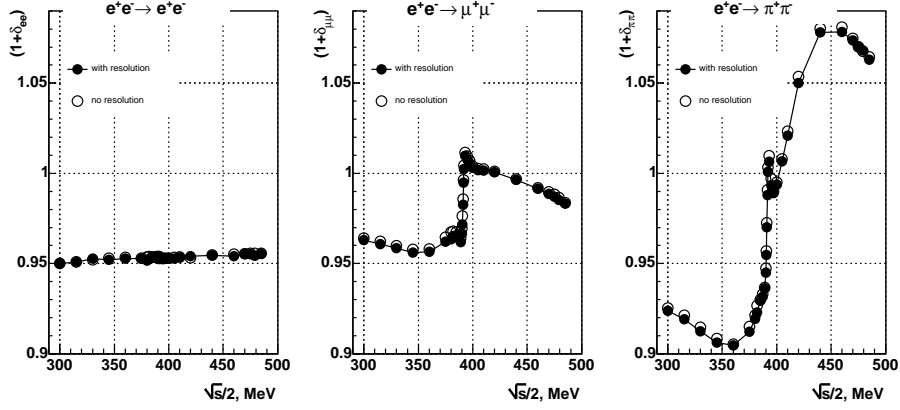}
\end{center}
\caption{\label{fig:rc} Radiative corrections for $\Theta_{min}=1.1$,
  $|\Delta\Theta|<0.25$, $|\Delta\varphi|<0.15$. Solid points and line
  represent the results of the calculations with the detector resolution taken
  into account. Circles represent the results of the  calculations with
  the ``ideal'' detector.}
\end{figure}

The results of the calculation are shown in Fig.\ \ref{fig:rc}. It is
clear that the contribution from the detector resolution 
is negligible for the standard data selection. It becomes much more
important if stricter cuts are applied.

\begin{figure}
\begin{center}
\begin{tabular}{cc}
\subfigure[\label{fig:dfpi_dtheta} Dependence of the form factor on
  $\Delta\Theta$ cut]
{\includegraphics[width=0.49\textwidth]{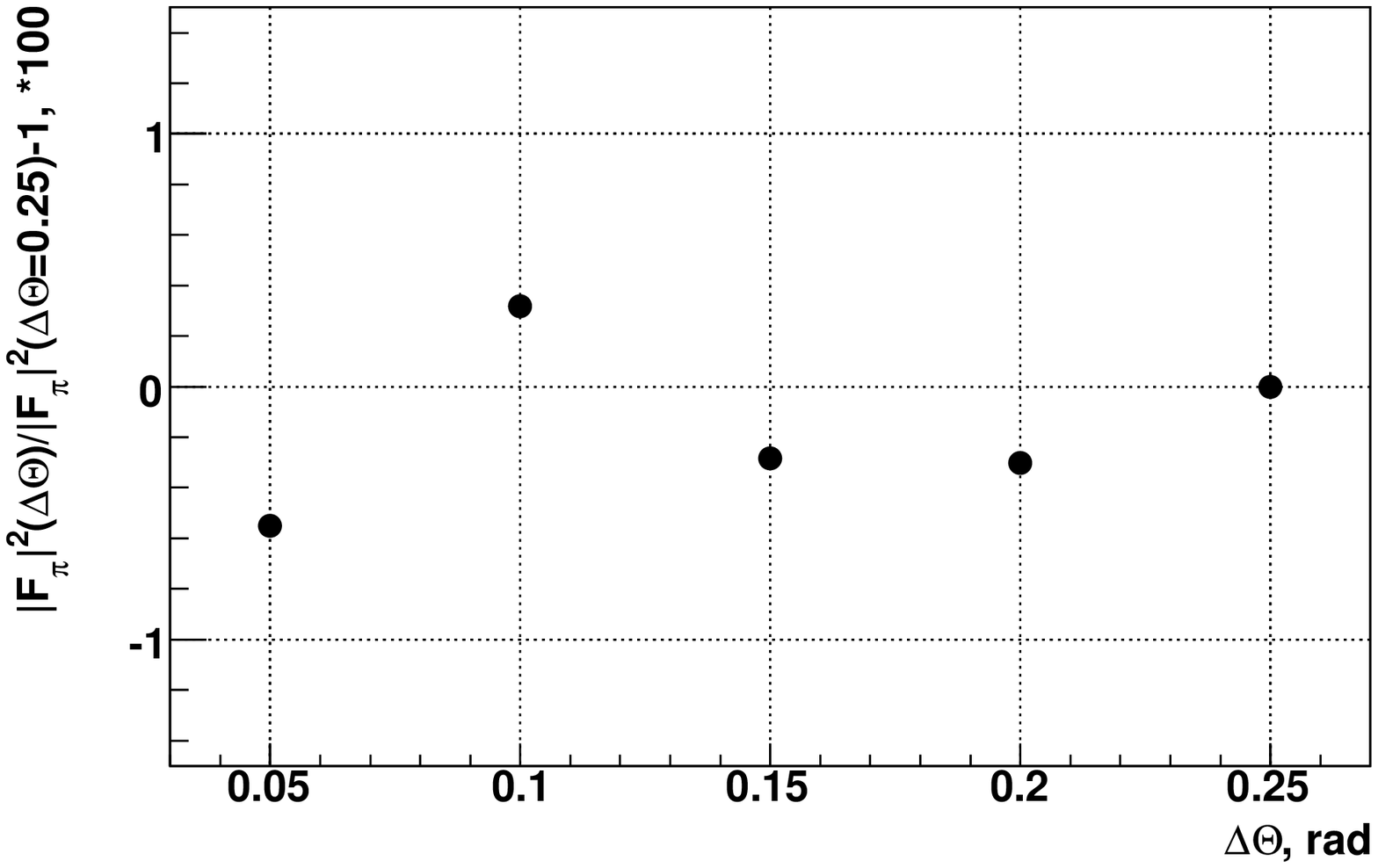}} &
\subfigure[\label{fig:dfpi_theta} Dependence of form factor on
  $\Theta_{min}$ cut (fiducial volume)]
{\includegraphics[width=0.49\textwidth]{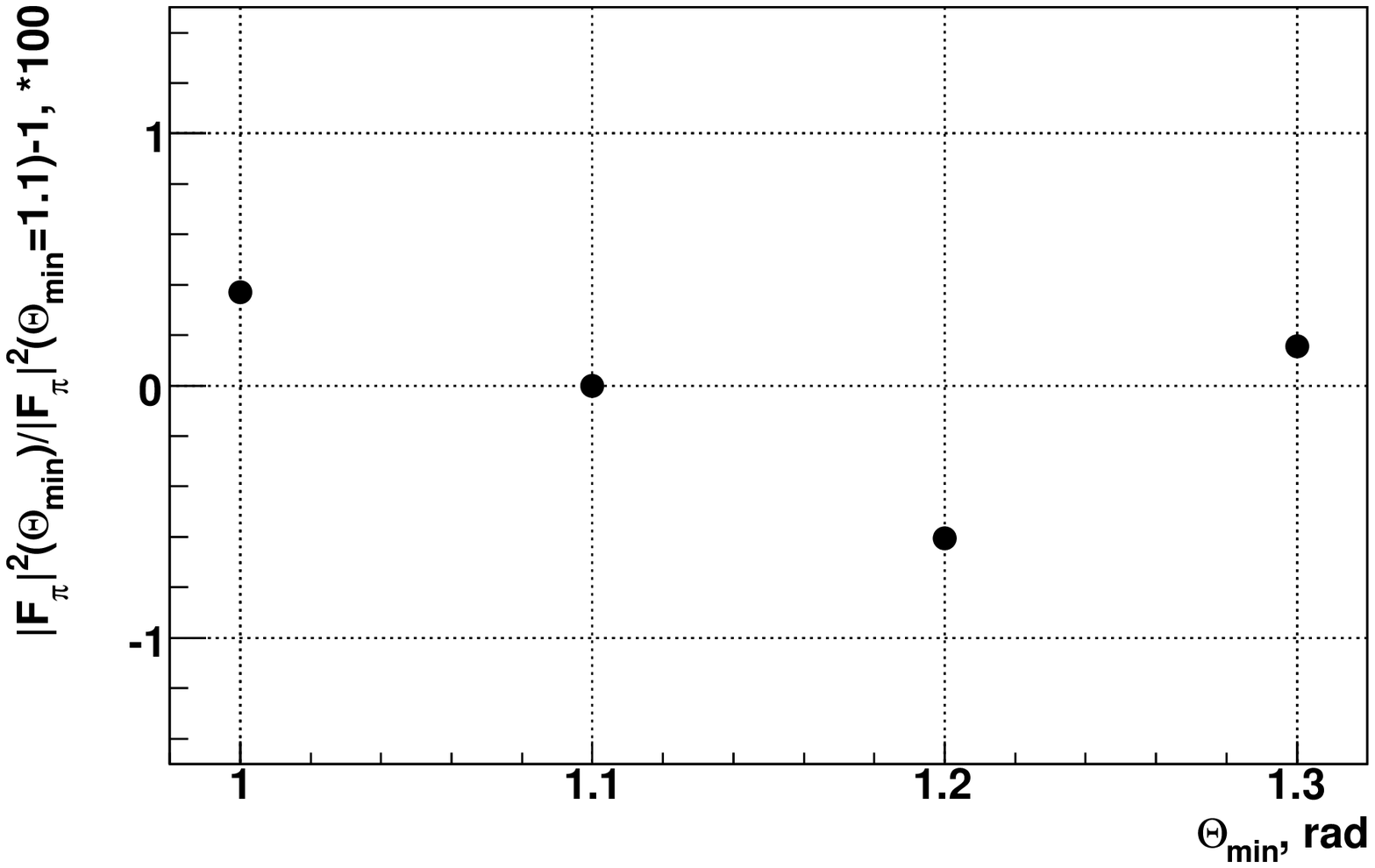}} 
\end{tabular}
\end{center}
\caption{\label{fig:fpivscuts} Difference between the results of the
  form factor measurements performed with different selection cuts.}
\end{figure}

As an indirect test of calculations, the whole data analysis procedure
was repeated for different cuts on $\Delta\Theta$, $\Delta\varphi$ and
$\Theta_{min}$. This test probes all pieces of the data analysis
procedure. But since these cuts affect the radiation corrections much
stronger than any other contribution, such as the efficiency, the
procedure mainly tests the radiative corrections. The results are
shown in Fig.\ \ref{fig:fpivscuts}. No changes outside the allowed limits
were observed. 

\subsection{\label{sec:other}Other corrections}

The corrections for the pion decay in flight $\Delta_D$, for the 
nuclear interactions of pions with the material of the beam pipe and
the drift chamber $\Delta_N$ and for the $\ee\to 3\pi$ background
$\Delta_{3\pi}$ were calculated with the help of simulation. The values
of the corrections are the same as those used for the 94-95 data
analysis \cite{rhoart}.

\begin{figure}
\begin{center}
\includegraphics[width=0.49\textwidth]{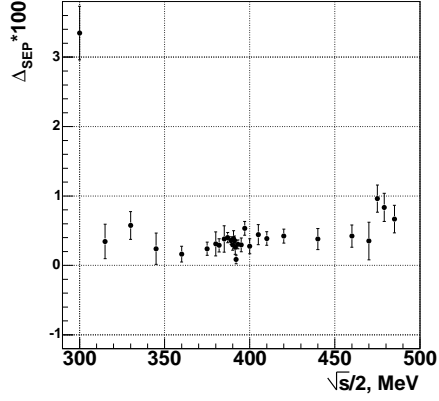}
\end{center}
\caption{\label{fig:sepcorr} Correction $\Delta_{sep}$ for the correlation
of the energy depositions of two particles due to the initial state radiation}
\end{figure}

One new correction, denoted as $\Delta_{sep}$ in (\ref{piform}), was applied
to the 98 data set. An event where one of the
original particles emits a hard photon is usually rejected
because the final particles are not back-to-back. But if both initial
particles radiate a hard photon, the final particles could stay
back-to-back and therefore be accepted for the analysis. The effect of
this double bremsstrahlung on the selection efficiency was taken into
account in the radiative correction calculation. But this effect also
introduces a correlation between the energy depositions of two final
particles in the calorimeter, which introduces a systematic shift of
the likelihood fit results.

Two approaches were used to take this effect into account. The first
one, applied in the analysis of the data above the $\varphi$-meson, is based on
the minimization of the modified likelihood function, where the
correlation term is introduced \cite{fedya}. The different approach
was used here. The correction to the results of the event separation
was evaluated using the Monte Carlo simulation and applied to the
final result. The size of the correction is small ($\approx$0.3\%), so
this simple approach does not introduce any sizable systematic
uncertainty. The correction is shown in Fig.\ \ref{fig:sepcorr}.
The same correction was evaluated when the double bremsstrahlung was switched
off in the Monte Carlo simulation. A significantly smaller effect was
observed, which proves that the double bremsstrahlung is the main
source of the shift. 

\subsection{Systematic errors}

\begin{table}
\begin{center}
\begin{tabular}{|lc|}
\hline
Source & Contribution, \% \\
\hline 
Event separation& 0.2 \\
Detection efficiency & 0.5 \\
Fiducial volume & 0.2 \\
Correction for pion losses & 0.2 \\
Beam energy determination & 0.3 \\
Radiative corrections & 0.4 \\
\hline
Total & 0.8 \\
\hline 
\end{tabular}
\end{center}
\caption{\label{tab:syst} Main sources of the systematic errors}
\end{table}

The main sources of the systematic error are summarized in Table
\ref{tab:syst}. Some contributions have not changed since the 94-95 data
analysis and are not discussed here, namely the fiducial volume and
the corrections for the pion losses.

The event separation was tested with the help of simulation. In these studies
the double bremsstrahlung was switched off in the primary generator, as was
discussed in section \ref{sec:other}. We studied how the following
contributions affect the results: the calorimeter calibration, the
initial and final state radiation,  ``dead'' crystals in the calorimeter.
The largest observed shift was about 0.2\%, which was taken as a
systematic error estimate. The additional double bremsstrahlung correction
is small, so it was assumed that it gives no contribution to the systematic
error. 

The systematic error of the efficiency measurement was discussed in
detail in section \ref{sec:eff}. We estimate this contribution to the
systematic error to be less than 0.5\%. It should be noted
that the difference between the reconstruction efficiencies for \eeee and
\eepp  events on average is $\approx$2\% and never exceeds
5\%. Therefore, the estimated 0.5\% suggests that this difference is
known to about 25\%. 

The absolute beam energy was determined from the value of the
collider magnetic field, which provided an accuracy better than
$\Delta E/E<10^{-3}$. The energy uncertainty leads to a 0.3\%
systematic error of the contribution to the $a^{(\rm had,LO)}_{\mu}$,
which we include in the total systematic error in Table \ref{tab:syst}.
The absolute energy scale can be calibrated
with the measurement of the mass of $\omega$-meson, the only narrow
resonance in the energy range under analysis. To do the calibration we
performed a fit of the measured form factor in which the $\omega$ mass
was a free fit parameter, and obtained 
$\momg(\mathrm{these\,data})-\momg(\mathrm{PDG2006}) = (0.4\pm 0.3)$ 
MeV, or $\Delta E/E \approx (5\pm 4)\cdot10^{-4}$. An
independent determination of the $\omega$ mass with the same data set
was performed in $e^+e^-\to\pi^0\gamma$ channel \cite{pi0g}. The
result $\momg(\pi^0\gamma)-\momg(\mathrm{PDG2006}) = (0.55 \pm 0.24)$
MeV, is consistent with our measurement.

The contribution of the radiative corrections to the systematic error 
is determined
by the precision of the ratio $(1+\delta_{\pi\pi})/(1+\delta_{ee})$. The
radiative correction to each final state is known to 0.2\% or better. We've
added the contributions from two final states linearly to obtain 0.4\% 
as the total contribution. Taking the detector resolution into account 
does not change the
results significantly, therefore no additional
contribution to the systematic error was added. 

\section{Results}

\begin{table}
\begin{center}
\begin{tabular}{|c|r@{$\,\pm\,$}l|r@{$\,\pm\,$}l|}
\hline
$\sqrt{s}$, MeV &\multicolumn{2}{|c|}{$|F_{\pi}|^2$} & \multicolumn{2}{|c|}{$\sigma^0_{\pi\pi(\gamma)}$, nb} \\\hline
600 &   7.89 &   0.33 &  330.1 &   13.7\\ 
630 &  10.53 &   0.29 &  415.6 &   11.5\\ 
660 &  14.21 &   0.34 &  529.1 &   12.7\\ 
690 &  21.27 &   0.42 &  746.4 &   14.7\\ 
720 &  31.96 &   0.41 & 1053.6 &   13.5\\ 
750 &  42.13 &   0.62 & 1296.6 &   19.0\\ 
760 &  43.62 &   0.83 & 1311.4 &   25.0\\ 
764 &  44.48 &   0.73 & 1325.8 &   21.7\\ 
770 &  44.17 &   0.74 & 1302.0 &   21.7\\ 
774 &  45.46 &   0.53 & 1332.2 &   15.4\\ 
778 &  44.52 &   0.42 & 1296.0 &   12.2\\ 
780 &  43.00 &   0.56 & 1237.5 &   16.0\\ 
781 &  41.40 &   0.53 & 1178.3 &   15.2\\ 
782 &  39.64 &   0.20 & 1111.7 &    5.6\\ 
783 &  36.46 &   0.42 & 1007.0 &   11.7\\ 
784 &  33.80 &   0.37 &  922.4 &   10.1\\ 
786 &  31.25 &   0.38 &  844.1 &   10.2\\ 
790 &  31.05 &   0.61 &  836.6 &   16.4\\ 
794 &  30.61 &   0.48 &  822.5 &   12.8\\ 
800 &  29.82 &   0.39 &  794.8 &   10.4\\ 
810 &  26.78 &   0.38 &  701.6 &    9.9\\ 
820 &  23.44 &   0.26 &  602.9 &    6.6\\ 
840 &  17.89 &   0.20 &  443.3 &    5.1\\ 
880 &  10.37 &   0.15 &  239.0 &    3.4\\ 
920 &   6.76 &   0.09 &  145.4 &    2.0\\ 
940 &   5.20 &   0.12 &  108.0 &    2.5\\ 
950 &   4.75 &   0.10 &   97.2 &    2.1\\ 
958 &   4.26 &   0.09 &   86.0 &    1.8\\ 
970 &   3.94 &   0.09 &   78.0 &    1.8\\ 
\hline
\end{tabular}

\end{center}
\caption{\label{table:fpi}Results of the pion form factor measurement
  based on the CMD-2 1998 data. Only statistical errors are shown.}
\end{table}

\begin{figure}
\begin{center}
\includegraphics[width=\textwidth]{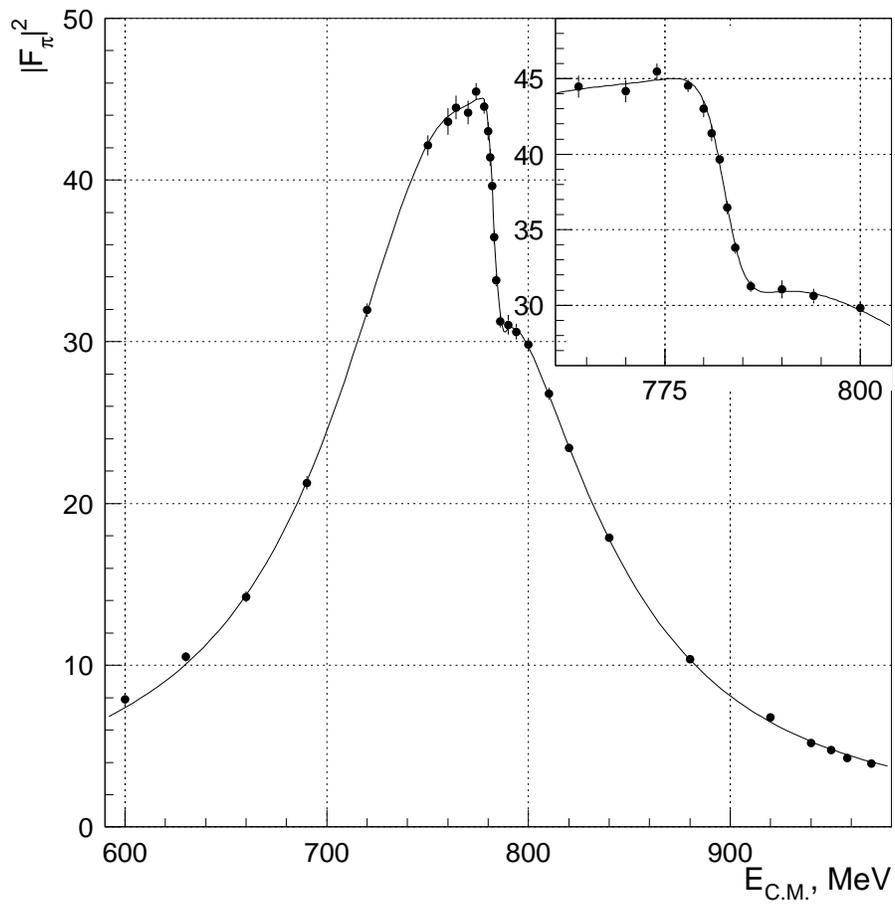}
\end{center}
\caption{\label{fig:rho98} Fit of the pion form factor measured in
  this work}
\end{figure}

\begin{figure}
\begin{center}
\begin{tabular}{cc}
\subfigure[\label{fig:fpicomp} Comparison of the pion formactor
  measured in this work with the previous measurements by CMD-2 and
  SND. The values are shown relative to the fit of the CMD-2 1998
  data.]
{\includegraphics[width=0.49\textwidth]{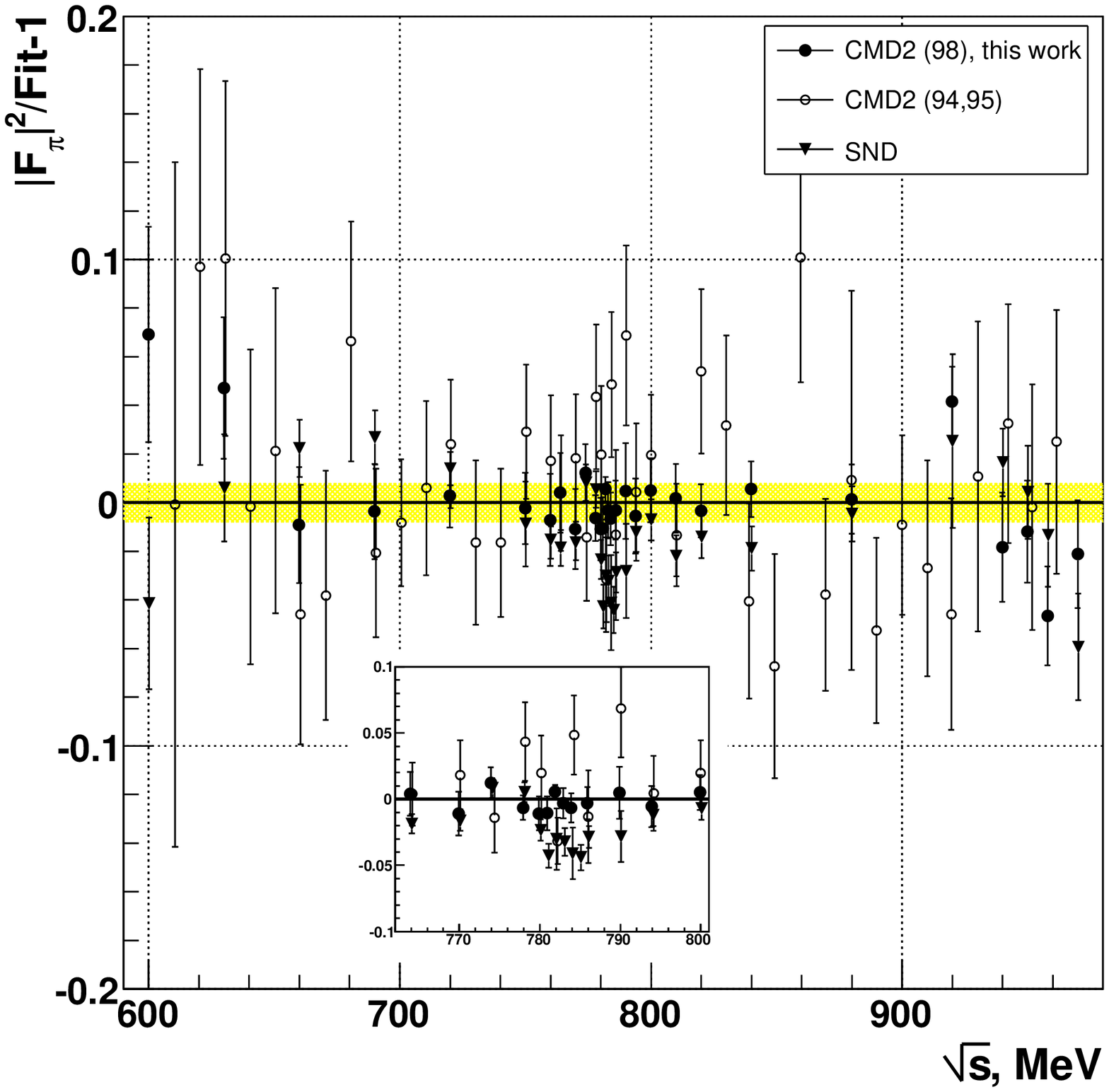}} &
\subfigure[\label{fig:kloe} Comparison of the pion formactor
  measured in this work with the KLOE measurement. The filled area
  shows the statistical error and the lines show the systematic error
  of the KLOE data.]
{\includegraphics[width=0.49\textwidth]{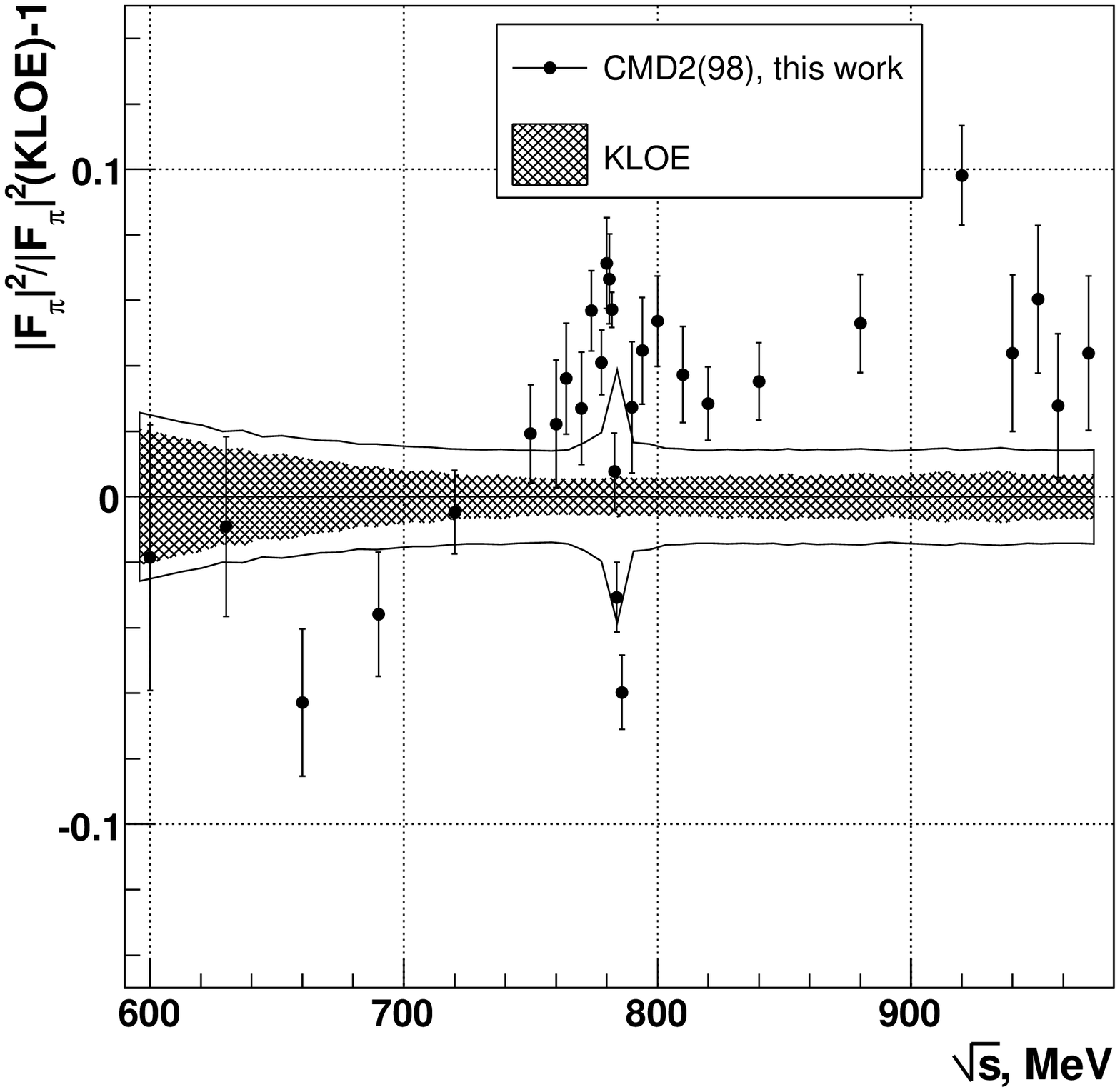}} 
\end{tabular}
\end{center}
\caption{Comparison of the pion formfactor measured in this work with
  other measurements}
\end{figure}

The measured values of the pion form factor are shown
in Table \ref{table:fpi}. Only the statistical errors are shown. Also
presented are the values of the bare 
$\eepp(\gamma)$ cross-section defined as
\[
\sigma^0_{\pi\pi(\gamma)} =
\frac{\pi\alpha^2}{3s}\beta^3_\pi \left| F_\pi(s)\right|^2
\cdot |1-\Pi(s)|^2
\cdot \left( 1+\frac{\alpha}{\pi}\Lambda(s) \right),
\]
where the factor $|1-\Pi(s)|^2$ excludes the effect of leptonic and
hadronic vacuum polarization and the factor $\Lambda(s)$ provides a
correction for the final state radiation.

To obtain the parameters of the $\rho(770)$ meson, the measured form factor
was fitted with the same model, as was used in our previous measurement,
which includes the contributions of the 
$\rho(770)$, $\omega(782)$ and $\rho(1450)$ and is based on the
Gounaris-Sakurai parameterization of the $\rho$ meson:
\[
F_\pi(s)=\frac{\GS_{\rho(770)}(s)\cdot
\left( {\mathstrut 1+\delta e^{-i\Phi_{\rho\omega}} \,
\frac{\displaystyle s}{\displaystyle M_\omega^2}} \, \BW_{\omega}(s) \right)
+ \beta \cdot \GS_{\rho(1450)}(s)
}{1+\beta}.
\]
It is assumed that $\omega$ decays to $2\pi$ through $\rho-\omega$
mixing only. More details on this model and the values of
all fixed parameters can be found in \cite{rhoart}. The 
following parameters of the $\rho(770)$ and $\omega(782)$ were obtained:
\begin{equation*}
\begin{array}{ll}
\mrho  & = (775.97 \pm 0.46 \pm 0.70)~ \text{MeV},
\vphantom{\Bigl( \Bigr)} \\
\Gamma_\rho  & = (145.98\pm 0.75 \pm 0.50)~ \text{MeV},
\vphantom{\Bigl( \Bigr)} \\
\Gamma(\rho\rightarrow e^+e^-)  & = (7.048\pm 0.057 \pm 0.050)~ \text{keV},
\vphantom{\Bigl( \Bigr)} \\
\mathcal{B}(\omega\rightarrow\pi^+\pi^-) & = ( 1.46\pm 0.12 \pm 0.02) \% ,
\vphantom{\Bigl( \Bigr)} \\
\Phi_{\rho\omega} & = 10.4^{\circ} \pm 1.6^{\circ} \pm 3.5^{\circ},
\vphantom{\Bigl( \Bigr)} \\
\beta & = -0.0859 \pm 0.0030 \pm 0.0027. \\
\vphantom{\Bigl( \Bigr)}
\end{array}
\end{equation*}
The first error is statistical and the second is systematic
taking into account the systematic uncertainties of the
data and the beam energy. These results are in good agreement with our
previous measurement \cite{rhoart,rhoerrata}. It should be noted that in our
parameterization $\beta$ represents the combined effect of the
$\rho(1450)$ and
$\rho(1700)$ and therefore cannot be used to obtain the $\rho(1450)\to
2\pi$ branching ratio. The value of
$\mathcal{B}(\omega\rightarrow\pi^+\pi^-)$ is calculated from $\delta$
assuming VDM relations and $\Gamma_{\omega ee}=(0.595 \pm 0.017)$
keV, as described in \cite{rhoart}.

Comparison between the results of this, our previous and the recently
published SND measurement \cite{sndfpi,snderrata} is shown in
Fig.\ \ref{fig:fpicomp}. The average difference between the two CMD-2
results is $(0.4\% \pm 0.6\% \pm 0.8\%)$, while between the SND and
this measurement it is $(-1.2\% \pm 0.4\% \pm 1.5\%)$, where the first
error is statistical and the second is the uncorrelated systematic one.

Recently the KLOE collaboration published the first measurement of
the \eepp cross-section \cite{KLOE} based on the analysis of the
distribution of the invariant mass of two pions in the
$\eepp+\gamma_{ISR}$ final state (the initial state radiation
or ISR approach). 
Comparison between the results of KLOE and our measurements
is shown in Fig.\ \ref{fig:kloe}. Only the statistical errors are
shown. There is some systematic difference between the results,
particularly in the energy range $\sqrt{s}>0.8$ GeV. Its nature is not 
understood at the moment. 

\begin{table}
\begin{center}
\begin{tabular}{|l|c|}
\hline
Experiment & $a^{\pi\pi,\rm LO}_{\mu}, 10^{-10}$ \\
\hline
CMD-2, 1994-1995 data & $362.1 \pm 2.4 \pm 2.2$ \\
\hline
CMD-2, 1998 data (this work) & $361.5 \pm 1.7 \pm 2.9$ \\
\hline
SND              & $361.0 \pm 1.2 \pm 4.7$ \\
\hline
KLOE & $357.2 \pm 0.5 \pm 4.6$ \\
\hline
\end{tabular}
\end{center}
\caption{\label{tab:amu}Comparison of the $2\pi$ contributions to the 
  $a^{(\rm had,LO)}_{\mu}$ coming from the energy range 
  $630<\sqrt{s}<958$ MeV. The first error is statistical and the
  second is systematic.} 
\end{table}

In Table~\ref{tab:amu} we compare the  $2\pi$
contributions of the above four measurements to the  leading order 
hadronic term of the muon anomalous magnetic moment
coming from the energy range 630--958~MeV.
The results are obtained by
direct integration of the experimental points using a trapezoidal method.
All four measurements give consistent values of $a^{\pi\pi,{\rm LO}}$.
In accordance with the discussion above, the CMD-2 result based on this
measurement (1998 data) has better statistical error than the
result based on our previous study. The combined uncertainty of the new
measurement is about the same as before.

This work is partially supported by the Russian Foundation for Basic
Research, grants 03-02-16477, 03-02-16280, 04-02-16217,
04-02-16223, 04-02-16434, 05-02-17169, 06-02-16156, 06-02-16445, 06-02-26859
and the U.S.\ National Science Foundation.

\end{document}